\documentclass{emulateapj}
\usepackage{amssymb}
\begin{document}

\title{The kinetic Sunyaev-Zel'dovitch effect as a dark energy probe}
\author{Simon DeDeo\altaffilmark{1,2}, David N. Spergel\altaffilmark{1}, Hy Trac\altaffilmark{1}}

\altaffiltext{1}{Department of Astrophysical Sciences, Princeton University, Princeton, New Jersey 08544, USA}
\altaffiltext{2}{\tt simon@astro.princeton.edu, simon@kicp.uchicago.edu}

\begin{abstract}
Upcoming observatories will be able to detect  the kinetic Sunyaev-Zel'dovitch (kSZ) effect with unprecendented signal-to-noise, and cross-correlations with foreground signals such as galaxy counts are a promising way to extract additional cosmological information. We consider how well a tomographic galaxy-count cross-correlation experiment, using data from WMAP, ACT and SALT, can significantly constrain the properties of dark energy. We include the need to model a wide range of effects, including those associated with complicated baryonic physics, in our analysis. We demonstrate how much of the cosmological information contained in the kSZ comes from larger scales than that in the galaxy power spectrum, and thus how use of the kSZ can help avoid difficult systematics associated with non-linear and scale-dependent bias at $k>1h$~Mpc$^{-1}$.
\end{abstract}

\maketitle

\section{Introduction}

Future microwave experiments such as \emph{Planck}, the \emph{Atacama Cosmology Telescope} (ACT) and the \emph{South Pole Telescope} (SPT) will map the cosmic microwave background (CMB) with unprecedented resolution and precision. In anticipation of this new era, much attention has been drawn to what can be learned from studies of not only the so-called ``primary'' CMB anisotropies -- those arising at the surface of last scattering -- but also ``secondary'' anisotropies, created by structures at much lower redshift.

Secondary anisotropies arise at all angular scales; the largest secondary anisotropy at the arcminute scales  probed by the experiments above is the thermal Sunyaev-Zel'dovitch (tSZ) effect~\citep{sz70}, caused by scattering of CMB photons from the hot, ionized gas associated with clusters. The tSZ has a unique, non-thermal signature which, in principle, allows it to be isolated from other contributions. 

After removal of the tSZ, the dominant source of arcminute anisotropy is expected to arise from the kinetic Sunyaev-Zel'dovich (kSZ) effect, caused by bulk flows in the same ionized gas, and the subject of this paper. It was first studied by~\cite{ov86} and for this reason is also known as the Ostriker-Vishniac effect (OV)\footnote{Because \cite{ov86} assumed both the velocity and density fields were linear, studies often refer to the ``OV'' effect as that arising from purely linear fluctuations, reserving the term ``kSZ'' for calculations that include a non-linear treatment of the density field.}. Other important contributors to arcminute-scale anisotropies include the gravitational lensing of primary anisotropies and, potentially, patchy reionization~\citep[\emph{e.g.}]{s03}; beyond $\ell$ of approximately $3000$, the amplitude of the primary anisotropies themselves is subdominant because of Silk damping~\citep{s68}.

In linear perturbation theory, the magnitude of velocity perturbations, and thus the kSZ, is directly related to both the growth rate of structure and the Hubble constant,
\begin{equation}
\label{velocity}
\vec{v}(\vec{k})=i\frac{d\ln{D}}{d\ln{a}}\frac{aH\delta(\vec{k})\vec{k}}{k^2},
\end{equation}
where $D$ is the growth factor at late times. A measurement of the amplitude of the kSZ and, in particular, a measurement of the amplitude as a function of redshift, would provide a new way to probe the nature of dark energy, which influences both the geometry of spacetime and the growth-rate of structure.

In this paper, we consider how well a measurement of the kSZ, in cross-correlation with a redshift-binned galaxy count signal such as might be provided by the upcoming \emph{Southern African Large Telescope} (SALT), might constrain properties of dark energy, including both the equation of state parameter $w$, and its derivative with respect to scale factor, $dw/da$. 

The kSZ is not a ``clean'' signal: it arises in part from the non-linear regime of structure formation, and depends on the details of baryonic physics on megaparsec scales. We first review, in Secs.~\ref{ksz-open} and~\ref{ksz}, the calculation of the angular power spectrum from the underlying three-dimensional density fields. 

In Secs.~\ref{hybrid} and~\ref{baryon}, we examine some of the approximations made in this calculation, comparing our results with those from simulations that include gas dynamics, and we introduce additional parameters into our model to take into account the need to model the effects of baryonic physics. This is an extension of the work of~\cite{d04}, who presented the first semi-analytic calculation of a cross-correlation between the kSZ and a density tracer (in that case, the weak lensing of galaxy images.) 

In Sec.~\ref{fisher}, we describe the Fisher matrix analysis we use to determine how the combination of ACT and SALT data, supplemented by large angular-scale data from the \emph{Wilkinson Microwave Anisotropy Probe} (WMAP), can determine cosmological parameters including $w$ and $dw/da$. We present our results in Sec.~\ref{all}. In Sec.~\ref{discardsection}, we examine how the use of the kSZ effect may allow us to partially circumvent difficult questions of small-scale physics. We discuss our results, and suggest future avenues for investigation, in Sec.~\ref{discussion}.

\section{The kSZ and Galaxy Count Samples}
\label{ksz-open}

In order to compute both the ACT kSZ signal and its cross-correlation with a SALT galaxy-count sample, we follow and extend the methods of~\cite{d04}. 

The kSZ signal is caused by the scattering of CMB photons off of moving ionized gas; unlike the tSZ it does not lead to a spectral distortion but rather only, to first order in $v/c$, a temperature increment or decrement depending on whether the halo is moving towards or away from the observer. The fractional change in temperature, $\Theta=\Delta T/T_0$, is
\begin{displaymath}
\Theta(\hat{\gamma})=-\int^{\eta_0}_0 d\eta g(\eta)\hat{\gamma}\cdot\vec{q}(\hat{\gamma}\eta,\eta),
\end{displaymath}
where $\hat{\gamma}$ is the unit vector pointing away from the observer and the momentum density, $q$, is defined as
\begin{displaymath}
\vec{q}(\hat{\gamma}\eta,\eta)=[1+\delta(\hat{\gamma}\eta)]\vec{v}(\hat{\gamma}\eta,\eta),
\end{displaymath}
$g$, the visibility function, is defined as
\begin{displaymath}
g(\eta)=x_e\tau_H(1+z)^2e^{-\tau},
\end{displaymath}
and $\tau_H$, the Thompson-scattering optical depth to the Hubble distance today, is
\begin{displaymath}
\tau_H=0.0691(1-Y_p)\Omega_bh,
\end{displaymath}
where $Y_p$ is the primordial Helium fraction. As discussed in Sec.~\ref{baryon} below, we describe the reionization $x_e$ as a linear function of redshift, zero before some early $z_{\mathrm{ri}}$.

For our fiducial SALT galaxy sample, we assume an idealized tracer with a linear bias on all scales. As discussed below, it is certainly possible to conduct an analysis with a less ideal tracer -- and some methods of kSZ-galaxy cross-correlation require detailed attention to biasing -- but for simplicity, and in ignorance of the actual biases that will be uncovered in a SALT sample, we do not consider this in detail.

We take our sample to have $6\times10^5$ galaxies over a total area of $100$ square degrees, with a median redshift of $0.7$, distributed log-normally with a first-moment equal to $0.3$ and with sufficient redshift information to allow the sample to be divided for the purposes of tomography into ten redshift bins with equal numbers of galaxies.

We define $W_{\delta,i}(\eta)$ as the window, or galaxy selection function of our sample; the subscript $i$ refers to one of ten equal number bins; $W$ is proportional to the number density of galaxies within the bin, and zero outside. For conciseness, we will omit this subscript in our later equations.

For the ACT experiment, we assume a $\theta_{\mathrm{FWHM}}$ of 1.7 arcminutes and noise $\sigma_{\mathrm{pix}}$ of 2.0 $\mu$K covering the same sky region as the SALT sample.

\section{Computing Angular Power Spectra}
\label{ksz}

The kSZ effect only becomes significant at very small angular scales ($\ell\sim10^3$). We thus use the Limber (Kaiser) approximation when projecting the various $3$-d density fields responsible for both the kSZ and galaxy angular power spectra. 

Our goal here is not to produce a ``final answer'' to the question of modelling the kSZ, but rather to demonstrate some useful approximations. We point out and model, to a first approximation, where difficult questions concerning gas physics arise. We demonstrate a model that roughly represents the amplitude and shape of the kSZ and its cross-correlation, and the sensitivity of these measurements to various cosmological parameters. One important question that we do not address here is the possibility of patchy reionization.

The kSZ effect is symmetric: an electron overdensity responsible for scattering a CMB photon is as likely to be moving towards us as away, and the temperature perturbation is proportional to the velocity. We again follow~\cite{d04} in correlating the \emph{square} of the (filtered) CMB temperature with the galaxy count signal. Filtering is necessary, since squaring in real space amounts to a convolution in Fourier space; the kSZ signal will be lost under the contribution from primary anisotropies and system noise unless the latter components are filtered:
\begin{displaymath}
\Theta_f(\vec{l})=f(\vec{l})\Theta(\vec{l}).
\end{displaymath}
The square of the filtered temperature field can then be written
\begin{displaymath}
\Theta^2_f(\vec{l})=\int\frac{d^2\vec{l}^\prime}{(2\pi)^2}\Theta_f(\vec{l}^\prime)\Theta_f(\vec{l}-\vec{l}^\prime)
\end{displaymath}
and the correlation between a galaxy-count and kSZ squared signal is
\begin{displaymath}
\langle\Theta^2_f(\vec{l})\delta(\vec{l}^\prime)\rangle=\int\frac{d^2\vec{l}^{\prime\prime}}{(2\pi)^2}\langle\Theta_f(\vec{l}^{\prime\prime})\Theta_f(\vec{l}-\vec{l}^{\prime\prime})\delta(\vec{l}^\prime)\rangle.
\end{displaymath}
The three-point function in the integral can be written as the projection of a three-dimensional power spectrum as
\begin{eqnarray*}
\langle\Theta_f(\vec{l})\Theta_f(\vec{l}^\prime)\delta(\vec{l}^{\prime\prime})\rangle & = & (2\pi)^2f(l^\prime)f(|\vec{l}-\vec{l}^{\prime\prime}|)\int d\eta\frac{W_\delta(\eta)}{\eta^2} \\
& & \times \int d\eta^\prime \frac{g(\eta^\prime)}{\eta^{\prime2}}\int d\eta^{\prime\prime}\frac{g(\eta^{\prime\prime})}{\eta^{\prime\prime}} \\
& & \times\int \frac{dk_z}{2\pi}e^{ik_z(\eta-\eta^\prime)}\int\frac{dk_z^\prime}{2\pi}e^{ik_z^{\prime}(\eta^{\prime\prime}-\eta^\prime)} \\
& & \times \delta_D\left(\frac{\vec{l^\prime}}{\eta}+\frac{\vec{l^{\prime\prime}}}{\eta^{\prime\prime}}+\frac{\vec{l}-\vec{l^{\prime\prime}}}{\eta^\prime}\right) \\
& & \times B_{q_{\hat{\gamma}}q_{\hat{\gamma}}\delta}\left[\frac{\vec{l}^{\prime\prime}}{\eta^{\prime\prime}}+k_z^\prime,\frac{\vec{l}-\vec{l}^{\prime\prime}}{\eta^\prime}-(k_z+k_z^\prime), \right.\\
& & \left.\frac{\vec{l}^\prime}{\eta}+k_z \right],
\end{eqnarray*}
where we define the hybrid bispectrum as
\begin{eqnarray*}
\langle q_{\hat{\gamma}}(\vec{k}_1)q_{\hat{\gamma}}(\vec{k}_2)\delta(\vec{k}_3)\rangle & = & (2\pi)^3B_{q_{\hat{\gamma}}q_{\hat{\gamma}}\delta}(\vec{k}_1,\vec{k}_2,\vec{k}_3) \\
& & \times\delta_D(\vec{k}_1+\vec{k}_2+\vec{k}_3),
\end{eqnarray*}
where $q_{\hat{\gamma}}$ is shorthand for $\vec{q}\cdot\hat{\gamma}$. Following \cite{bkj00}, we ignore radial modes (the Limber approximation), taking $k_z$ to be much less than $l/\eta$ and finding
\begin{eqnarray*}
\langle\Theta_f(\vec{l})\Theta_f(\vec{l}^\prime)\delta(\vec{l}^{\prime\prime})\rangle & \approx & (2\pi)^2\delta_D(\vec{l}+\vec{l}^\prime)f(l^{\prime\prime})f(|\vec{l}-\vec{l}^{\prime\prime}|) \\
& & \times\int\frac{d\eta}{\eta^4} W_\delta(\eta)[g(\eta)]^2 \\
& & \times B_{q_{\hat{\gamma}}q_{\hat{\gamma}}\delta}\left(\frac{\vec{l}^{\prime\prime}}{\eta^{\prime\prime}},\frac{\vec{l}-\vec{l}^{\prime\prime}}{\eta^\prime},\frac{\vec{l}^\prime}{\eta}\right).\\
\end{eqnarray*}
This gives us all we need to compute $C^{\delta\Theta_f^2}_l$, defined as
\begin{displaymath}
C^{\delta\Theta_f^2}_l=\int\frac{d\eta}{\eta^2}W_\delta(\eta)[g(\eta)]^2T\left(k=\frac{l}{\eta},\eta\right),
\end{displaymath}
where $T$, the ``triangle power spectrum'', is defined as
\begin{displaymath}
T(k,\eta)=\int\frac{d^2\vec{q}}{(2\pi)^2}f(q\eta)f(|\vec{k}+\vec{q}|\eta)B_{q_{\hat{\gamma}}q_{\hat{\gamma}}\delta}(\vec{q},-\vec{k}-\vec{q},\vec{k}).
\end{displaymath}
As discussed in \cite{d04}, this can be understood as an integral over all triangles with sides $(\vec{q},-\vec{k}-\vec{q},\vec{k})$, where the triangles are restricted to lie on planes of constant redshift. It is a way to ``collapse down'' the rich set of information contained in the three-point function.

When considering future analyses of the kSZ signal dealing with real data, it seems likely that a gain in signal-to-noise can be had by constructing a filter to preferentially magnify those triangle configurations sensitive to the parameters of interest, in a fashion similar to analyses that have been done in the area of  lensing of the primary anisotropies~\citep{cht00}. Since our goal in this paper is to get a rough estimate of the overall signal-to-noise as it pertains to extraction of cosmological parameters, we ignore this subtlety for now.

\section{The Hybrid Bispectrum}
\label{hybrid}

The momentum perturbation is $[1+\delta(\vec{x})]\vec{v}(\vec{x})$. In Fourier space, this is
\begin{equation}
\label{momentum}
\vec{q}(\vec{k})=\vec{v}(\vec{k})+\int\frac{d^3k^\prime}{(2\pi)^3}\vec{v}(\vec{k}^\prime)\delta(\vec{k}-\vec{k}^\prime).
\end{equation}
We are interested in the component parallel to the line of sight, $\vec{q}_{\hat{\gamma}}$. We will work exclusively in the small angle approximation, so that a position on the sky is written $\vec{\theta}$. The kSZ autocorrelation in the two-point Limber (Kaiser) approximation has been worked out in many places; we will work out the kSZ-kSZ-galaxy correlation in the three-point Limber approximation.

Determining an analytic form for the kSZ signal is a series of approximations. We will refrain from substituting in definite forms for the various two- and three-point functions until the next section. As discussed in the original analysis~\citep{ov86}, the ``trick'' is to note that, in the Limber approximation, only $\vec{q}$ modes with $\vec{k}$ perpendicular to the line of sight contribute. We can write down the ``curl'' component, $\vec{q}_p$, which has velocity perpendicular to its wavenumber. The magnitude of the line-of-sight momentum component squared is, finally, $(1/2)\vec{q}_p\cdot\vec{q}_p$, where the factor of one-half comes from the averaging over $\sin^2\phi$; $\vec{q}_p$ can point anywhere in the plane perpendicular to the line of sight.

Within linear theory velocities are purely gradient and the first term in Eq.~\ref{momentum} will not contribute to the curl component, so that only the mode coupling term $\delta\vec{v}$ is important in each instance of $\vec{q_{\hat{\gamma}}}$. Many recent analyses have taken the density field to be non-linear, but kept the assumption of linear and gradient velocity fields~\citep{h00, m02}.

\cite{z04}, however, have raised doubts as to whether the assumption of a linear velocity gives the full amplitude of the kSZ signal. In particular, they find that, at scales relevant to ACT, curl modes of the non-linear velocity field may be of similar magnitude to the gradient modes estimated with the linear density field. The velocity curl mode is generated by shell crossings and the resultant gas physics, and is probably intractable analytically. While still considering only the $\delta\vec{v}$ term, they suggest a phenomenological substitution of the fully non-linear density field into the equation for the velocity from linear perturbations (Eq.~\ref{velocity}.) Although this substitution is hard to justify physically, it accords very well with their simulations, and we follow their practice.

Apart from this substitution, \cite{z04} use the same reasoning as \cite{m02} when computing the kSZ power analytically, except that in the latter paper certain approximations are later introduced to make the mathematics of the two-point function more tractable. As we show below, when computing the three-point function there are ten separate terms to keep track of; introducing the analogous approximation to that made in \cite{m02} significantly simplifies the calculation while still providing a good estimate of the true kSZ signal at small scales.

A potentially complicating factor has been raised by \cite{p05}, who noted that a naive application of theory underpredicts even large-scale velocity correlations estimated when the kSZ effect is used to determine the proper motions of clusters: in that case, the fact that clusters form preferentially in already overdense regions demands that biasing effects in the velocity must be taken into account. 

In this paper, by contrast, we consider for simplicity a cross-correlation of the kSZ with an ``ideal'' galaxy sample with linear bias, and do not preferentially select clusters.

We now proceed to calculate the three-dimensional three-point function for the kSZ-kSZ-galaxy count signal,
\begin{displaymath}
\langle q_{\hat{\gamma}}(\vec{k}_1)q_{\hat{\gamma}}(\vec{k}_2)\delta(\vec{k}_3)\rangle = \frac{1}{2}\langle\vec{q}_p(\vec{k}_1)\cdot\vec{q}_p(\vec{k}_2) \delta(\vec{k}_3)\rangle. 
\end{displaymath}
There will be $5\times4/2!$, \emph{i.e.}, ten, different products of two- and three-point functionals to consider, but many of these will turn out to be zero. Schematically, and assuming that the connected five-point term is zero, the terms will look like
\begin{eqnarray}
\label{schematic}
\langle\delta v\delta v \delta\rangle & = & \langle\delta\delta\rangle\langle v v  \delta\rangle
+\langle\delta v\rangle\langle \delta v  \delta\rangle
+\langle v v\rangle\langle \delta \delta  \delta\rangle \nonumber \\
& & +\langle\delta  \delta\rangle\langle \delta v  \delta\rangle
+\langle v  \delta\rangle\langle \delta v  \delta\rangle,
\end{eqnarray}
with some terms occuring more than once, with a permutation of the (suppressed) indicies. The \emph{Ansatz} of~\cite{d04} effectively considered only the third term of the series. Below, we will judge the validity of this approximation. Let us, to begin with, examine this third term to determine its complete analytic form. We have
\begin{eqnarray*}
\label{dore}
\langle vv\rangle\langle \delta\delta\delta\rangle & = & (\delta_D^{ij}-\hat{k}_1^i\hat{k}_2^j)\int\frac{d^3k}{(2\pi)^3}\int\frac{d^3k^\prime}{(2\pi)^3} \langle v^i(\vec{k}) v^j(\vec{k}^\prime)\rangle \\
& & \langle \delta(\vec{k}_1-\vec{k})\delta(\vec{k}_2-\vec{k}^\prime)\delta(\vec{k}_3)\rangle.
\end{eqnarray*}

Substituting in the velocity power spectrum and the density bispectrum we find this to be equal to
\begin{eqnarray*}
& & (\delta_D^{ij}-\hat{k}_1^i\hat{k}_2^j)\int\frac{d^3k}{(2\pi)^3}\int\frac{d^3k^\prime}{(2\pi)^3} (2\pi)^3\delta_D(\vec{k}+\vec{k}^\prime)\hat{k}^i\hat{k^{\prime j}}P_{vv}(\vec{k}) \\
& & (2\pi)^3\delta_D(\vec{k}_1+\vec{k}_2+\vec{k}_3-\vec{k}-\vec{k}^\prime)B_{\delta\delta\delta}(\vec{k}_1-\vec{k},\vec{k}_2-\vec{k}^\prime,\vec{k}_3),
\end{eqnarray*}
which, upon working out the delta functions, becomes
\begin{eqnarray*}
& & (2\pi)^3\delta_D(\vec{k}_1+\vec{k}_2+\vec{k}_3)\int\frac{d^3k}{(2\pi)^3}[1-(\hat{k}_1\cdot \hat{k})(\hat{k}_2\cdot \hat{k})]
\\ & & P_{vv}(\vec{k})B_{\delta\delta\delta}(\vec{k}_1-\vec{k},\vec{k}_2+\vec{k},\vec{k}_3).
\end{eqnarray*}

We now examine the high-$k_{1,2,3}$, nonlinear behaviour. Because the peak of the integrand will occur at some small $\vec{k}$, we can, following~\cite{m02}, drop terms of order $k/k_{1,2,3}$. After averaging over the $k$ orientation, and projecting along the line of sight, and assuming the other terms in Eq.~\ref{schematic} are negligible, we find the hybrid bispectrum, $B_{q_{\hat{\gamma}}q_{\hat{\gamma}}\delta}(\vec{k}_1,\vec{k}_2,\vec{k}_3)$, to be
\begin{equation} 
\label{b}
 \frac{1}{2}\left(1-\frac{1}{3}\hat{k}_1\cdot\hat{k}_2\right)
v^2_{{\rm rms}}B_{\delta\delta\delta}(\vec{k}_1,\vec{k}_2,\vec{k}_3).
\end{equation}
This analytic result is very close to the \emph{Ansatz}, made in~\cite{d04}, Eq.~30, which turns out to be a slight underestimate of our Eq.~\ref{b}.

We have a number of other terms to consider, now, to see if they can be eliminated in the high-$k$ approximation. The essential ``trick'' is that there are geometric factors in front of these other terms that either cancel or are of order $k/k_{1,2,3}$. Fig.~1 shows the ten different cases. As can be seen, the term that leads to Eq.~\ref{b} is the leading-order term in the nonlinear regime. A more precise model can be constructed by retaining terms of order $k/k_{1,2,3}$ in Eq.~\ref{b}, and including the three additional terms, listed in Fig.~1, that do not cancel by angular symmetry.

\begin{figure}
\label{cases}
\begin{center}
\leavevmode
\begin{tabular}{c|c}
term & scaling \\
\hline
$\langle v^i(\vec{k}) v^j(\vec{k}^\prime)\rangle\langle \delta(\vec{k}_1-\vec{k})\delta(\vec{k}_2-\vec{k}^\prime)\delta(\vec{k}_3)\rangle$ & 1 \\
$\langle v^i(\vec{k}) \delta(\vec{k}_1-\vec{k}) \rangle\langle v^j(\vec{k}^\prime)\delta(\vec{k}_2-\vec{k}^\prime)\delta(\vec{k}_3)\rangle$ & 0 \\
$\langle v^i(\vec{k})\delta(\vec{k}_2-\vec{k}^\prime)\rangle\langle \delta(\vec{k}_1-\vec{k})v^j(\vec{k}^\prime)\delta(\vec{k}_3)\rangle$ & $k/k_2$ \\
$\langle \delta(\vec{k}_2-\vec{k}^\prime) v^j(\vec{k}^\prime)\rangle\langle \delta(\vec{k}_1-\vec{k})v^i(\vec{k})\delta(\vec{k}_3)\rangle$ & 0 \\
$\langle\delta(\vec{k}_1-\vec{k}) v^j(\vec{k}^\prime)\rangle\langle  v^i(\vec{k})\delta(\vec{k}_2-\vec{k}^\prime)\delta(\vec{k}_3)\rangle$ & $k/k_1$ \\

$\langle v^i(\vec{k}) \delta(\vec{k}_3)\rangle\langle \delta(\vec{k}_1-\vec{k})\delta(\vec{k}_2-\vec{k}^\prime) v^j(\vec{k}^\prime)\rangle$ & 0 \\
$\langle \delta(\vec{k}_3) v^j(\vec{k}^\prime)\rangle\langle \delta(\vec{k}_1-\vec{k})\delta(\vec{k}_2-\vec{k}^\prime)v^i(\vec{k})\rangle$ & 0 \\

$\langle\delta(\vec{k}_1-\vec{k})\delta(\vec{k}_2-\vec{k}^\prime) \rangle\langle v^i(\vec{k}) v^j(\vec{k}^\prime)\delta(\vec{k}_3)\rangle$ & $k/k_3$ \\
$\langle \delta(\vec{k}_2-\vec{k}^\prime)\delta(\vec{k}_3)\rangle\langle v^i(\vec{k}) v^j(\vec{k}^\prime) \delta(\vec{k}_1-\vec{k})\rangle$ & 0 \\
$\langle\delta(\vec{k}_1-\vec{k})\delta(\vec{k}_3)\rangle\langle  v^i(\vec{k}) v^j(\vec{k}^\prime)\delta(\vec{k}_2-\vec{k}^\prime)\rangle$ & 0 \\
\hline
\end{tabular}
\end{center}
\caption{The ten terms of the the kSZ-kSZ-galaxy cross-correlation, and the scaling of their associated geometric coefficients after integrating over $\vec{k}^\prime$. A zero implies that the term cancels after integrating over the angular direction of $\vec{k}$.}
\end{figure}
 
Having reduced the hybrid bispectrum down to a formula involving the density bispectrum, we use the fitting formula of~\cite{sc01} to determine the (non-linear) $B_{\delta\delta\delta}$; schematically, this involves permuted pairs of the non-linear power spectrum, $P^{\mathrm{nl}}$:
\begin{displaymath}
B_{\delta\delta\delta}(\vec{k}_1,\vec{k}_2,\vec{k}_3)=2F^{\mathrm{eff}}_2(\vec{k}_1,\vec{k}_2)P^{\mathrm{nl}}_{\delta\delta}(k_1)P^{\mathrm{nl}}_{\delta\delta}(k_2)+\mathrm{cyclic}.
\end{displaymath}
To compute the non-linear power spectrum, we use the \emph{halofit} code, which implements the formulae of \cite{sp03}. We modify the code to allow for a $w(a)$ dependence. Note that \emph{halofit} uses an analytic approximation to the transfer function that does not include the phenomenon of ``baryon wiggles''; our estimate of the cosmological parameters thus does not include the additional information contained in these features.
 
The function $F_2$ appearing above is a complicated, though computable, function of the parameters estimated by \emph{halofit}. As described in~\cite{d04}, integrals involving $B_{\delta\delta\delta}$ may be significantly simplified by invoking the same, high-$k$, approximation used to derive Eq.~\ref{b}. The triangle power spectrum, $T$, defined above may then be written as
\begin{displaymath}
\Delta_T^2(k,z)=\frac{1}{2}v_{\mathrm{rms}}^2\Delta_{\mathrm{nl}}^2(k,z)E_3(k,z)
\end{displaymath}
where the dimensionless power spectra $\Delta_T^2$ and $\Delta_{\mathrm{nl}}^2$ are defined as
\begin{eqnarray*}
\Delta_T^2=\frac{k^3}{2\pi^2}T & ; & \Delta_{\mathrm{nl}}^2=\frac{k^3}{2\pi^2}P^{\mathrm{nl}}_{\delta\delta}
\end{eqnarray*}
and $E_3$ is defined as in~\cite{d04} based on the expressions found in~\cite{sc01}.

\section{Modeling Baryonic Physics}
\label{baryon}

As hot gas condenses into stars, is influenced by supernova and AGN feedback, or even as it simply responds to its own pressure gradients, we expect the distribution of the baryons responsible for the kSZ effect to diverge from that of the underlying dark matter. Current X-ray observations~\citep{v05} and cluster simulations~\citep{k05} find that the baryon fraction, $f_{\mathrm{gas}}$, approaches the cosmological value beyond $r_{500}$, the radius at which the cluster is $500$ times overdense compared to cosmic, but the details of this behaviour, and how it depends on cluster mass, have yet to be completely understood.

We are thus unable to precisely model the systematics of baryon physics as they relate to the determination of cosmological parameters. However, it seems likely that a combination of computer simulations and observational work will gradually refine our understanding of baryon physics to the point where systematic errors may be strongly constrained. In anticipation of that era, we include a ``toy'' model of baryon physics whose free parameters are included in the final Fisher matrix analysis.

Our model first allows the total baryon fraction to evolve as a function of scale factor; \emph{i.e.}, we take $f_{\mathrm{gas}}(a)$ to be $\Omega_b(1+ca)$, where $a$ is the scale factor and $c$ is a free parameter to be determined by the model fit. This is intended to model the incorporation of baryons into stars. 

Since the \emph{halofit} code transfer function does not include the ``baryon wiggles,'' a changing $\Omega_b$ at late redshifts is indistinguishable from a changing ionization fraction, $x_e$. From the combination of the WMAP measurement of the optical depth to reionization, and the detection of the Gunn-Peterson trough at $z$ of 6.28 \citep{b01}, we know that the reionization history of the universe is more complex than a simple step function at some early redshift; the inclusion of the parameter $c$ goes some way to including the need to ``fit out'' for these effects when determining the cosmological parameters.

While we have thus included the possibility of a time-dependent $x_e$, there is also the possibility of a spatial dependence, a phenomenon known as patchy reionization and already detected at low redshifts~\citep{o05}. Patchy reionization has the potential to reduce the signal-to-noise of the kSZ detection; studies of its effect on kSZ detection suggest that ``worst case'' models may reduce the signal-to-noise by a factor of two. For simplicity, we do not consider the effect here.

Secondly, we allow for a  ``smoothing scale'', $\sigma_{\mathrm{sm}}$, the width of a Gaussian convolved with the dark matter distribution to determine the baryon distribution. In accordance with both observation and simulation, the baryons are ``smoothed'' out compared with the dark matter. The smoothing scale not only introduces partial degeneracies in our final parameters, it also reduces the amplitude of the kSZ signal, and thus our overall signal-to-noise.

The inclusion of a smoothing scale fits both with simple adiabatic gas simulations, where pressure leads to gas profiles less concentrated than the associated dark matter, and with more complicated simulations that include cooling and star formation, which produce evidence compatible with smoothing~\citep{k05}, including a greatly reduced $f_{\mathrm{gas}}$ near the cluster center. Observations~\citep{v05} also find similar effects compatible with our model.

\begin{figure}
\includegraphics[width=3.375in]{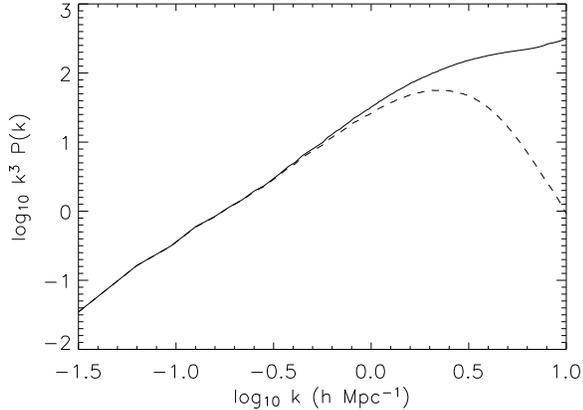}
\caption{A comparison of the baryon (dashed line) and dark matter (undashed line) power spectra at redshift zero from our simulations. At scales smaller than approximately $7$ Mpc the two spectra begin to diverge; our smoothing prescription (described in the text) allows us to accurately predict the baryon power spectrum from that of the dark matter down to scales of approximately $1.5$ Mpc.}
\label{smoothing}
\end{figure}

To determine a realistic value of $\sigma_{\mathrm{sm}}$, we analyse a $200 h^{-1}$~Mpc hydrodynamic simulation. The Eulerian TVD+PM code of~\cite{t04} was run with $1024^3$ grid cells and $512^3$ dark matter particles.  The cell spacing of  $195 h^{-1}$~kpc and particle mass of $4.96\times10^9 h^{-1} M\odot$ provides sufficient resolution for this work. Note that this code has been used previously in similar studies of the SZ effect~\citep[\emph{e.g.}]{z04}. Results for a $100 h^{-1}$~Mpc simulation with the same number of cells and particles are identical in the region of overlap.

In Fig.~\ref{smoothing} we plot the baryon and dark matter power spectra from this simulation at redshift zero, when we expect effects from baryon physics to be most pronounced. As can be seen, the baryon power spectrum diverges from the dark matter at lengths of around $7$ Mpc. 

Setting the smoothing scale, $\sigma_{\mathrm{sm}}$, equivalent to a length of $2.2$ Mpc, we can model the baryon power spectrum to ten percent accuracy down to lengths around $1.5$ Mpc at redshift zero. Below this length our naive model tends to \emph{underpredict} structure in the baryon distribution. Our simulations suggest that $\sigma_{\mathrm{sm}}$, if taken to be a comoving quantity, changes little over the redshift range of interest.

We do not expect our parameters to be the ``final answer'' to the difficult problems of modeling baryonic physics. However, given the current uncertainty, our two parameters, $c$ and $\sigma_{\mathrm{sm}}$, covering the possibility of redshift and scale dependence of the ionized baryon distribution, seem to be a reasonable way to study both how uncertainties about the effects of baryon physics degrade our ability to study the properties of dark energy, and how studies of the kSZ effect may be a new means, complimentary to studies in the X-ray, to probe the distribution of gas.

\section{Determining Cosmological Parameters}
\label{fisher}

We combine the following information to determine our set of parameters: large angular-scale WMAP data, small angular-scale ACT data, and a SALT galaxy-count sample that overlaps the ACT strip. We consider a ``tomographic'' analysis: \emph{i.e.}, in our analysis we divide the galaxy count sample into ten redshift bins with equal numbers of galaxies in each. Given the log-normal galaxy distribution described above in Sec.~\ref{ksz-open}, we are most sensitive to fluctuations around redshift of $0.7$.

As mentioned above, we require a filter, $f(l)$, to filter the temperature map to avoid contamination of the kSZ signal from ``out of band'' power upon squaring. As in~\cite{d04}, we use the filter that maximizes the kSZ signal alone,
\begin{equation}
\label{filter}
f(l)=\frac{C^{\mathrm{kSZ}}_l}{C^{N}_l+C^{\mathrm{kSZ}}_l},
\end{equation}
where the ``noise'' term includes both instrumental noise and any sources of CMB power not associated with the kSZ. 


To characterize the performance of an ideal experiment, we compute the Fisher matrix errors; from the Crameo-Rao inequality, these are the best possible error bars that can be achieved by the above described experiments given their noise characteristics and assuming Gaussianity. The Fisher matrix is
\begin{equation}
F_{\alpha\beta}=\sum_l\sum_{X,Y}\frac{\partial C^{X}_l}{\partial \lambda_\alpha}\mathrm{Cov}\left[C^{X}_lC^{Y}_l\right]^{-1}\frac{\partial C^{Y}_l}{\partial \lambda_\beta}
\end{equation}
where $\lambda_\alpha$ refers to the set of parameters we hope to determine, and $X$ and $Y$ refer to the twenty-one different measurements, \emph{i.e.}, temperature--temperature, galaxy--galaxy (ten bins), and temperature-squared--galaxy (ten bins.) In the case of the temperature--temperature correlation, we use WMAP data on large scales, where the sky coverage beats down cosmic variance, and ACT for small scales.

The covariance matrix has off-diagonal terms; in particular, the off-diagonal term
\begin{equation}
\label{crossy}
\mathrm{Cov}\left[C^{\mathrm{G-G}}_lC^{\mathrm{T^2-G}}_l\right]=\frac{2C^{\mathrm{T^2-G}}_l(C^{\mathrm{G-G}}_l+C^{\mathrm{N}}_l)}{(2l+1)f_{\mathrm{sky}}}
\end{equation}
is non-zero. Inclusion of this term is necessary to avoid ``double-counting'' information in the final analysis (a related subtlety is that the filter defined in Eq.~\ref{filter} should not be varied from the fiducial value when computing the derivatives $dC_l/d\lambda$.) An additional term in Eq.~\ref{crossy}, proportional to the three-point correlation of the galaxy signal multiplied by the total temperature RMS, we have neglected as being strongly subdominant.

Other off-diagonal terms are zero, at least in the (good) approximation that the (unsquared) temperature--galaxy correlation is zero; we expect the lensing signal to be much smaller than the kSZ. Since the SALT galaxy sample only covers 100 deg$^2$, we can ignore the (very) large-scale correlations induced by the integral Sachs-Wolfe effect which were detected in a cross-correlation of WMAP and SDSS data. Since the covariance matrix is block diagonal, its inversion to produce the Fisher matrix is then trivial.

\section{All-Scale Limits}
\label{all}

To demonstrate for which parameters a kSZ-galaxy cross-correlation provides the greatest information, we give the errors for two separate analyses, one that uses WMAP and ACT temperature anisotropy, and SALT galaxy count, data (set ``A''), and one that uses this data plus the cross-correlation signal between the ACT and SALT data (set ``B''). 

We first consider constraints that can be made by including information on all scales -- \emph{i.e.}, when we use all available $\ell$ measurements limited only by Poisson and instrument noise. Our results are listed in the table in Fig.~3, which gives the fiducial cosmological model and estimated errors on our suite of parameters. 

For simplicity, we have assumed a flat universe prior. Because of recent interest in dark energy models that display ``phantom'' behaviour, we do not require $w\geq-1$; however, we assume that the dark energy fluid does not lead to appreciable clustering, as is the case, \emph{e.g.}, when the true sound speed $c_s$ is unity.

\begin{figure}
\begin{center}
\leavevmode
\begin{tabular}{c|c|c|c}
parameter & fiducial value & A error & B error \\
\hline $\Omega_m h^2$ & 0.1400 & 0.0016 & 0.0016 \\
$\Omega_b h^2$ & 0.02400 & 0.00019 & 0.00019 \\
$d_{\mathrm{LSS}}$ & 1.390 Gpc & 0.029 Gpc & 0.023 Gpc  \\
$\sigma_8$ & 0.84 & 0.10 & 0.054  \\
$w$ & -1.000 & 0.099 & 0.081 \\
$dw/da$ & 0.00 & 0.18 & 0.18 \\
``$c$'' & 0.00 & 1.1 & 0.48  \\
$\sigma_{\mathrm{sm}}$ & 0.350 Mpc$^{-1}$& 0.071 Mpc$^{-1}$ & 0.062 Mpc$^{-1}$ \\
$z_{\mathrm{ri}}$ & 17.00 & 0.18 & 0.17 \\
$b$ & 1.00 & 0.18 & 0.10 \\
\hline
\end{tabular}
\end{center}
\caption{Computed errors from a Fisher matrix analysis combining WMAP, ACT and SALT datasets as described. Set ``A'' contains the temperature anisotropy and galaxy count signals, set ``B'' includes the cross-correlation between the temperature squared and galaxy count signals.}
\label{errors}
\end{figure}

As can be seen, the addition of the cross-correlation information reduces the error on $\sigma_8$, $d_{\mathrm{LSS}}$ and $w$, and gives significant constraints on the parameters associated with late-time gas physics, $c$ and $\sigma_{\mathrm{sm}}$.

The sources of these results are not hard to understand. The strong $\sigma_8$ dependence in the cross-correlation signal has already been noted by \cite{d04}, and the reduction in error on $d_{\mathrm{LSS}}$ and $w$ are presumably due to the dependence of the velocity field on the instantaneous value of the Hubble parameter, Eq.~\ref{velocity}. 

That errors on $dw/da$ are not significantly improved is most likely due to our choice of parametrization; a dark energy model that restricted variations in $w$ to very late times ($z<1$) would presumably receive much better constraints from the addition of the cross-correlation information. The redshift of reionization, $z_{\mathrm{ri}}$, is strongly constrained by the overall amplitude of the kSZ signal, as noted by \cite{z04}; a slight improvement in limits on including the cross-correlation comes from removing degeneracies with other parameters, in particular $\sigma_8$. Meanwhile, the significant improvement in parameters $c$ and $\sigma_{\mathrm{sm}}$ is unsurprising; cross-correlation is an excellent method to study the late-time time evolution and spatial dependence of the ionized baryons.

\section{Linear vs. Non-linear Scales}
\label{discardsection}

While we have attempted to address a collection of different effects that complicate the analysis of galaxy and kSZ information, there remain a number of difficulties that will complicate any future analysis. A significant and troublesome systematic effect, that comes in at small scales, is the emergence of non-linear, and possibly non-deterministic, bias in the galaxy count signal \citep{sw98}. Our analysis above neglected these effects.

As can be seen in Eq.~\ref{velocity}, the velocity signal is proportional to $\delta/k^{-2}$, and thus is stronger at larger scales where bias models are better constrained. It can thus be said that the kSZ signal, in addition to having different systematics from a galaxy-count signal, also provides ``higher-quality'' information that may be less susceptible to the difficulties inherent in understanding structures on very small scales.

We can quantify this suggestion by redoing the analysis in the previous section and discarding small-scale information. Since the bulk of our galaxies are found at $z\approx0.7$, this can be done simply by including only low-$\ell$ information. We thus consider a cut at $\ell\approx1750$, corresponding to a comoving scale of $1h$~Mpc$^{-1}$at $z$ of $0.7$. In general, effects such as a scale dependence in the bias are expected to become strong at this scale; it is important to note that such effects do appear to a lesser extent at even larger scales~\citep{b99}.

The table in Fig.~4 shows the results of this cut. As can be seen, when the signal is limited to larger scales, the kSZ signal indeed becomes more important in constraining cosmological parameters. A ``conservative'' analysis of the WMAP, ACT and SALT data that discards smaller scales will improve its errors by a factor of two or more when it includes information derived from a cross-correlation of the kSZ and galaxy count signals.

\begin{figure}
\label{discard}
\begin{center}
\leavevmode
\begin{tabular}{c|c|c}
parameter & $1h$~Mpc$^{-1}$A error &$1h$~Mpc$^{-1}$B error \\
\hline $\Omega_m h^2$ & 0.0021 & 0.0017  \\
$\Omega_b h^2$ & 0.00043 & 0.00020  \\
$d_{\mathrm{LSS}}$ & 0.032 Gpc & 0.029 Gpc \\
$\sigma_8$ & 0.55 & 0.22  \\
$w$ & 0.11 & 0.10 \\
$dw/da$ & 0.19 & 0.19 \\
``$c$'' & n.c. & 1.4 \\
$\sigma_{\mathrm{sm}}$ & 0.81 Mpc$^{-1}$& 0.14 Mpc$^{-1}$ \\
$z_{\mathrm{ri}}$ & 0.39 & 0.19 \\ 
$b$ & 0.72 & 0.31 \\
\hline
\end{tabular}
\end{center}
\caption{The effect of restricting the signal to large scales where the bias is more easily modeled. We truncate the signal at $1h$~Mpc$^{-1}$ and consider both the case in which the CMB and galaxy-count signals are used only separately (``set A''), and the ``set B'' analysis, where the cross-correlation information is included. The term ``n.c.'' is used when constraints are so loose as to signal a significantly non-Gaussian distribution where parameters refer to physical quantities that may not go negative.}
\end{figure}

As mentioned above, the bias is not expected to be completely scale-free linear at  any of the scales probed by an ACT/SALT experiment. Furthermore, even when we restrict our study to $k<1 h$~Mpc$^{-1}$, we are still in the ``non-linear'' regime in one important sense; below scales of $0.1 h$~Mpc$^{-1}$ (corresponding to $\ell$ of $175$), the dark matter power spectrum can no longer be well-approximated by linear theory alone. Any analysis of either the kSZ or galaxy signal must always take this kind of non-linearity into account.

\section{Discussion}
\label{discussion}

One of the lessons of our paper is that the kSZ is of interest not only for studies of dark energy, but also for investigations of the distribution and evolution of baryons in the universe. Our work is complementary to that of~\cite{j05}; both papers consider the possibilities of using the ACT measurement of the kSZ to study cosmological parameters, but do so in different ways. 

In particular, \cite{j05} uses the kSZ to provide an unbiased measurement of individual cluster motions, whereas we consider the possibility of a statistical extraction of velocity fields on a wide range of scales.  The physics behind the generation of peculiar velocities is rich, and a number of different kinds of analysis made be done to produce results with different systematics.

Both a galaxy-count and a galaxy-kSZ cross-correlation have excellent prospects for the measurement of cosmological parameters, but have different systematics and are sensitive to different physics. As we have discussed in Sec.~\ref{discardsection}, the kSZ can be said to provide ``high quality'' information, by relying for the bulk of its signal on velocity flows on large scales where a cross-correlation does not have to worry as much about the effects of non-linear and non-deterministic bias.

\section{Acknowledgments}

We thank Olivier Dor\'{e} and Raul Jimenez for sharing their insights during the progress of this research.

\bibliography{paper}

\end{document}